\documentclass[conference]{IEEEtran}
\IEEEoverridecommandlockouts
\usepackage{cite}
\usepackage{amsmath,amssymb,amsfonts}
\usepackage{algorithmicx}
\usepackage{graphicx}
\usepackage{textcomp}
\usepackage{xcolor}
\usepackage{algorithm}
\usepackage{algpseudocode} 
\usepackage{amsthm}
\usepackage{subfigure}

\usepackage{multirow}

\def\BibTeX{{\rm B\kern-.05em{\sc i\kern-.025em b}\kern-.08em
    T\kern-.1667em\lower.7ex\hbox{E}\kern-.125emX}}
\begin{document}

\title{SeqBalance: Congestion-Aware Load Balancing with no Reordering for RoCE\\
}

\author{

	\IEEEauthorblockN{Huimin Luo$^{1}$, Jiao Zhang$^{1,2,*}$, Mingxuan Yu$^1$, Yongchen Pan$^1$, Tian Pan$^{1,2}$, and Tao Huang$^{1,2}$}
	\IEEEauthorblockA{$^1$ State Key Laboratory of Networking and Switching Technology,
		BUPT, Beijing, China}
	\IEEEauthorblockA{$^2$ Purple Mountain Laboratories, Nanjing, China}

	Emails:luohuimin@bupt.edu.cn, jiaozhang@bupt.edu.cn, mingxuan\_yu@bupt.edu.cn, \\panyongchen@bupt.edu.cn, pan@bupt.edu.cn, htao@bupt.edu.cn
	\thanks{*The corresponding author of this work is Jiao Zhang.}
}
\maketitle

\begin{abstract}
Remote Direct Memory Access (RDMA) is widely used in data center networks because of its high performance. However, due to the characteristics of RDMA's retransmission strategy and the traffic mode of AI training, current load balancing schemes for data center networks are unsuitable for RDMA. In this paper, we propose SeqBalance, a load balancing framework designed for RDMA. SeqBalance implements fine-grained load balancing for RDMA through a reasonable design and does not cause reordering problems. SeqBalance's designs are all based on existing commercial RNICs and commercial programmable switches, so they are compatible with existing data center networks. We have implemented SeqBalance in Mellanox CX-6 RNICs and Tofino switches. The results of hardware testbed experiments and large-scale simulations show that compared with existing load balancing schemes, SeqBalance improves 18.7\% and 33.2\% on average FCT and $99^{th}$ percentile FCT.
\end{abstract}


\section{Introduction}
The escalating demands placed on applications within data centers are propelling the evolution of high-speed networks, which are expected to provide high throughput, low latency and low CPU overhead. For instance, the distributed machine learning training service \cite{b1} necessitates a bandwidth exceeding 100 Gbps, while Web Search \cite{b2} imposes stringent requirements on latency, demanding the swiftest possible query response time. RDMA facilitates direct data exchange in the host's memory by circumventing the kernel. Moreover, it offloads network I/O functions to a dedicated RDMA network interface card (RNIC) to mitigate CPU overhead. RDMA ensures remarkable performance metrics, featuring high throughput (100/400 Gbps) and ultra-low latency ($\sim$ 1 us), all achieved with low CPU overhead. Owing to these inherent advantages, RoCEv2 (RDMA on Converged Ethernet Version 2) is extensively employed in modern data center networks to attain superior performance \cite{b3, b4}.

The prevailing RDMA architecture relies on connection-based single-path transmission, wherein the flow is confined to a fixed path once the RDMA connection is established. However, contemporary data center topologies are typically designed with scalability in mind, featuring multiple end-to-end paths between any two server racks \cite{b5, b6}. The exclusive reliance on a single path in existing RDMA not only renders it susceptible to path failures but also poses challenges in fully harnessing the potential of multiple paths within modern data centers. For example, in distributed deep learning applications, RDMA transmission cannot fully utilize the link resources of the network \cite{b24}. Consequently, to maximize the utilization of link resources within the network, the implementation of a load balancing strategy for RDMA becomes imperative.

The prevalent load balancing scheme in contemporary data centers is Equal Cost Multi-Path (ECMP). ECMP, a per-hop-based load balancing solution, involves each node making a local decision to determine the path of the next hop. However, several researches indicate that ECMP exhibits suboptimal load balancing, particularly in scenarios of highly skewed traffic distribution \cite{b5, b7, b8,b9}.
This inadequacy stems from two primary factors. Firstly, ECMP employs random hashing of flows to different paths, and with the uneven distribution of traffic sizes in the data center, hash collisions arising from flows of disparate sizes result in significant load imbalances. Secondly, ECMP, relying on local decisions, lacks awareness of potential downstream congestion. Consequently, it performs suboptimally in asymmetric topologies, especially in the presence of link failures \cite{b10}. Especially in the AI training, the traffic mode dominated by elephant flow only has a maximum throughput of 60\% under ECMP \cite{b37}.

Considerable efforts have been dedicated to addressing the limitations of ECMP and enhancing load balancing in data centers \cite{b5, b6, b7, b8, b11, b12,b41}. The proposed solutions fall into two primary categories. The first category comprises load balancing schemes based on per-packet routing \cite{b11, b12}. While these schemes can achieve nearly optimal load balancing, they concurrently introduce the risk of generating out-of-order packets. The second category involves load balancing schemes based on flowlet routing \cite{b6, b8, b13, b14, b15}. These schemes segment traffic into multiple flowlets for transmission based on the inactive time threshold, effectively mitigating the occurrence of out-of-order packets. However, the suitability of these schemes depends on the specific traffic characteristics. It is noteworthy that neither of these two types of TCP-based load balancing schemes proves suitable for RDMA ($\S$\uppercase\expandafter{\romannumeral2}).

The implementation of load balancing in RDMA presents three significant challenges. \textbf{Firstly, determining the appropriate granularity} for load balancing is pivotal in RDMA. Load balancing schemes, such as those based on ECMP, may result in load imbalances due to the uneven distribution of large and small flows. Meanwhile, per-packet routing-based schemes introduce the risk of out-of-order packet delivery, severely impacting RDMA performance. Although flowlet routing schemes can avoid out-of-order delivery, identifying flowlets proves challenging in RDMA. \textbf{Secondly, how to avoid out-of-order delivery of packets} while performing fine-grained load balancing is also vital for RDMA. Given that RDMA employs the Go-back-N retransmission mode, any instance of out-of-order packet delivery significantly impairs performance. \textbf{Thirdly, compatibility with commercial RNICs} is crucial for the designed load balancing scheme. At present, some work on RDMA will use independently designed RNIC for design (such as FPGA-based scheme design and implementation) due to some resource problems (such as small cache size) of modern commercial network cards \cite{b16,b17}, but this scheme is not compatible with the current commercial RNICs that have been deployed on a large scale in the data center network. Therefore, the designed load balancing scheme should be compatible with current commercial RNICs.

In this paper, we propose \textbf{SeqBalance} (or \textit{Sequential Balance}), a load balancing framework designed for RDMA. SeqBalance leverages \textit{SeqBalance Shaper} (Section $\S$\uppercase\expandafter{\romannumeral3}.C) to partition the RDMA flow into $N$ sub-flows of equal size for transmission, delegating the load balancing decision to the Top-of-Rack (ToR) switch. SeqBalance's entire load decision process occurs within the network, without necessitating alterations to the transport layer or the use of specialized RNICs, ensuring compatibility with existing data center network equipment. Simultaneously, SeqBalance achieves multi-path transmission through the sub-flows generated by \textit{SeqBalance Shaper}, and also avoids the problem of out-of-order packets in RDMA that seriously affects performance. The SeqBalance prototype is implemented using NVIDIA Mellanox ConnectX-6 RNICs \cite{b18} and the Intel Tofino programmable switch \cite{b19}. Large-scale simulations and hardware testbed results demonstrate that SeqBalance outperforms existing solutions.

The contributions of this paper are as follows:
\begin{itemize}
	\item We design \textit{SeqBalance Shaper} ($\S$\uppercase\expandafter{\romannumeral3}.C) to achieve fine-grained segmentation of RDMA flows without causing out-of-order packets. This module has been implemented on the RDMA driver.
	\item We design ($\S$\uppercase\expandafter{\romannumeral3}) and implement SeqBalance, a congestion-aware load balancing framework for RDMA. SeqBalance is easy to implement in hardware and is compatible with RNICs in existing data centers.
	\item We extensively evaluate SeqBalance with large-scale simulations and hardware testbed. The simulation and experimental results show that compared with the state-of-the-art, the average FCT and $99^{th}$ percentile FCT of SeqBalance improve by 18.7\% and 33.2\%, respectively.
\end{itemize}

The paper is organized as follows: we first introduce motivation and the design decisions in $\S$\uppercase\expandafter{\romannumeral2}. In $\S$\uppercase\expandafter{\romannumeral3}, we propose SeqBalance, a load balancing framework for RDMA. In $\S$\uppercase\expandafter{\romannumeral4}, we evaluate SeqBalance in hardware testbed and large-scale simulation. In $\S$\uppercase\expandafter{\romannumeral5}, we offer additional discussions. We summarize related work in $\S$\uppercase\expandafter{\romannumeral6}. We conclude the paper in in $\S$\uppercase\expandafter{\romannumeral7}.

\section{Motivation and Design Decisions}
\subsection{RDMA is not tolerant of out-of-order packets}

Unlike traditional TCP software-based transmission, RDMA represents a hardware-based approach that implements transmission functionality within network card hardware, offering kernel bypass and zero-copy interfaces for applications. Consequently, in contrast to TCP, RDMA attains superior throughput, reduced latency, and low CPU overhead, leveraging its kernel bypass and hardware forwarding advantages.

However, RDMA has its unique order-preserving mechanism that adapts to lossless network environment. The order-preserving of RDMA is based on the granularity of queue pair (QP), which means that RDMA performs ordered transmission at the granularity of an entire flow. In the Base Transport Header (BTH) of RDMA packets, there is a field called Packet Sequence Number (PSN). This field is specific to each QP connection. In each pair of QP connections, there is an independent PSN. When the packet passes through the IB transport layer (the layer is solidified in RNIC), the PSN code in the packet header will continue to accumulate. At the same time, when the packet is transmitted to the destination end RNIC, the PSN code is identified at the IB transport layer. If the PSN is not received in order (i.e., packets arrive out of order), it will immediately initiate packet loss retransmission, which seriously affects transmission performance.

\begin{table}[t]
	\begin{center}
		\caption{FCT for flows of different sizes when a packet is manually delayed.}
		\begin{tabular}{c|cc} 
			\multirow{2}{*}{\textbf{FCT normalized to no out-of-order packet}} & \multicolumn{2}{c}{\textbf{Message Size}}\\
			\cline{2-3}
			& \textbf{64KB} & \textbf{1MB} \\
			\hline
			average FCT & 5.77 & 3.01\\
			99$^{th}$-percentile FCT & 5.63 & 3.28\\
		\end{tabular}
	\end{center}
\end{table}

To illustrate this issue, we conducted experiments involving two sets of fixed-size flows, deliberately introducing a delay in the transmission of one packet, leading to the out-of-order arrival of a packet. The FCT sizes for both small flows (64KB) and large flows (10MB) are presented in Table \uppercase\expandafter{\romannumeral1}. Notably, the out-of-order arrival of a data packet results in a minimum threefold increase in FCT. Consequently, load balancing schemes for RDMA should prioritize avoiding out-of-order packet arrivals to ensure optimal transmission performance.

\subsection{The Traffic Mode of AI Training}

AI training requires high transmission rates, therefore, RDMA technology is often used as the communication method in AI training \cite{b36}. However, under the traffic load of AI, the load balancing effect of ECMP is not good. This is caused by the traffic characteristics of AI training. The flow generated by AI and storage workloads is often small, but very large. Therefore, the randomness is very low, and the load balancing effect using ECMP is poor \cite{b37}.

From the perspective of entropy theory, entropy in the field of networks can be defined as a measure of the randomness of data in network packets or flows. ECMP relies on high entropy to evenly distribute traffic across multiple links. The traffic model of traditional cloud network services generates thoughts of flows randomly connecting to various clients, providing a high entropy. However, in the traffic mode of AI, the number of flows is often small, while the size of flows is large. The low entropy traffic pattern of AI training is also known as the elephant traffic distribution. In the elephant flow scenario, the impact of hash conflicts is particularly significant, with only 60\% of the maximum throughput under ECMP \cite{b37}.

Therefore, in the AI training where RDMA transmission is most commonly used, load balancing algorithms are crucial.

\subsection{Design Decisions}
In SeqBalance, we make several design choices to build a load balancing framework that works with RDMA. We now discuss our design decisions.
\\[4pt]
\textbf{End-host vs In-Network Load Balancing Decision. }Determining where to implement load balancing stands as the primary consideration, specifically whether on the end-host or within the network. The nature of RDMA transfers, bypassing the kernel, precludes load balancing implementation in the transfer stack. Thus, if end-host load balancing is sought, it must be confined to operations on RNIC. However, it is difficult for end-host RNICs to perceive congestion: the ECN and RTT information measured by Programmable Congestion Control (PCC)\cite{b38} is only used for congestion control of the RNIC and cannot reach the upper layer for processing. MP-RDMA\cite{b16} achieves end-host multipath RDMA transmission through custom designed FPGA hardware, thereby implementing load balancing.
However, the prevalent use of commercial RNICs, such as Mellanox CX-6 \cite{b18}, in most data centers poses compatibility challenges with custom-designed RNICs like MP-RDMA. Moreover, the high construction cost in practical scenarios discourages their deployment. Consequently, our approach involves implementing load balancing decisions within the network. This strategic choice ensures SeqBalance's compatibility with RNICs commonly found in existing data centers, facilitating straightforward deployment. This also corresponds to our \textbf{Challenge \#3}.
\begin{figure}[t]
		\centering
		\begin{tabular}[t]{@{}r@{}l@{}}
			\subfigure[TCP] 
			{
				
				\centering          
				\includegraphics[width=4cm]{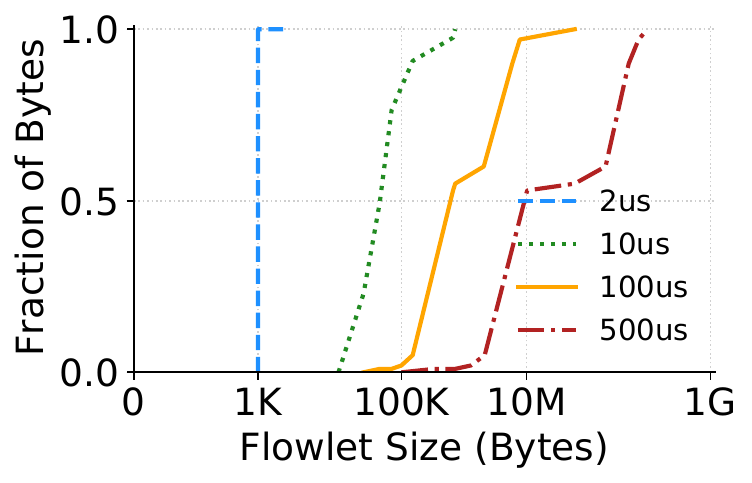}   
				
			}
			\subfigure[RDMA] 
			{
				
				\centering          
				\includegraphics[width=4cm]{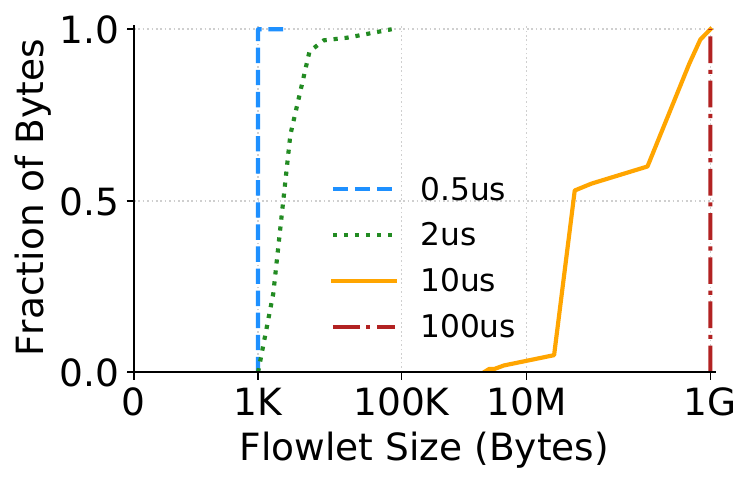}   
				
			}
		\end{tabular}
		\caption{Flowlet characteristics in TCP and RDMA.}
\end{figure}
\\[4pt]
\textbf{Load Balancing Granularity. }
As mentioned earlier, ECMP has low efficiency in AI scenarios where RDMA transmission is commonly used nowadays. Therefore, choosing the granularity of load balancing is important, which corresponds to our \textbf{Challenge \#1}.

One such alternative is the per-packet load balancing scheme \cite{b11,b12}. Despite the careful design of schemes like DRILL and Presto to minimize the generation of out-of-order packets, complete avoidance remains elusive, leading to subpar RDMA performance \cite{b34}. Consequently, per-packet rerouting schemes are deemed unsuitable for our considerations.

An alternative approach is the flowlet-based load balancing scheme \cite{b6,b14}. Flowlet distinguishes bursts of packets based on the inactive time interval. The inactive time interval exceeds the maximum delay difference between paths, allowing two bursts to traverse distinct paths without reordering packets. However, the characteristics of RDMA traffic render flowlet-based methods less applicable. This conclusion is drawn from experimental testing and observation. Fig. 1 depicts flowlet sizes for both TCP and RDMA traffic using varying inactive time thresholds. Even under a 10us inactive time threshold, RDMA flowlets are consistently larger than 10MB, and under a 100us threshold, flowlet sizes approach 1GB. In contrast, with a 100us threshold for TCP, flowlet sizes are predominantly below 10MB. Essentially, RDMA encounters greater challenges in identifying flowlets compared to TCP under the same inactive time threshold, aligning with findings in \cite{b16}. In practical scenarios, the inactive time threshold typically ranges between 100-500us \cite{b6}. This underscores the operational difficulty of identifying flowlets in RDMA traffic. The outcome is attributed to RDMA's lack of TCP-like burst transfers and its sustained high-rate traffic. Consequently, flowlet-based schemes are excluded from our considerations.

The determination of suitable load balancing granularity in RDMA prompts the question: what is the most effective choice? Drawing inspiration from the Justitia \cite{b21}, we opt for sub-flow as the load balancing granularity, striking a balance between per-packet and per-flow granularity. In this approach, the flow is partitioned into $N$ sub-flows, each hashed to distinct paths for achieving load balancing. See $\S$\uppercase\expandafter{\romannumeral3} for detailed design.
\\[4pt]
\textbf{Single-Path vs Multi-Path Transmission. }RDMA operates on a point-to-point transmission model. Upon establishing a QP connection, RDMA flow follows a predetermined path. 
In the case of substantial data transfers, RDMA exclusively employs single-path transfers. Consequently, if the established path encounters a failure, the RDMA transport lacks the inherent capability to automatically reroute traffic to an alternative path for transmission.

Distributed deep learning exemplifies a common scenario demanding extensive data transmission within data centers. For instance, Facebook employs Gloo \cite{b28}, a collective communication library, for model synchronization. The integration of RDMA transmission is a prevalent approach adopted to enhance the efficiency of large data transmission. In contemporary data center networks, the leaf-spine topology stands out as a widely utilized configuration, providing end-to-end path diversity between any server racks. Maestro\cite{b24} conducted experiments revealing that the application of distributed deep learning relying on point-to-point RDMA transmission falls short in fully leveraging the network's link resources. Meanwhile, the completion of LLM training is determined by the slowest GPU transmission, so imbalanced transmission can greatly affect training efficiency \cite{b37}. In order to increase entropy, introduce more randomness, and improve load balancing efficiency, we choose a multipath transmission scheme, which also corresponds to the sub-flow granularity we have chosen. MP-RDMA \cite{b16} stands out as a notable multi-path load balancing scheme tailored for RDMA. However, a notable drawback is its dependency on a custom RNIC implementation using FPGA, rendering it incompatible with widely deployed commercial RNICs. Consequently, this restricts its applicability in contemporary, mature, large-scale data centers. 

Therefore, in order to address \textbf{Challenge \#2} (avoiding packet out of order delivery) and \textbf{Challenge \#3}, we design SeqBalance, a multi-path load balancing scheme designed to seamlessly integrate with commercial RNICs, to solve the problem of sequential delivery during multipath transmission and achieve non out-of-order multipath transmission.
\\[4pt]
\textbf{Per-Hop vs End-to-End Multipathing. }The final design consideration revolves around whether load balancing should be based on a per-hop or end-to-end approach. In the case of SeqBalance, we opt to implement load balancing on an end-to-end basis within the network. This involves executing routing decisions on the ToR switch ($\S$\uppercase\expandafter{\romannumeral3}.B) and storing path information within the switch. Storing path information in this manner facilitates end-to-end path selection throughout the network. Notably, the stored information pertains only to paths experiencing congestion, ensuring minimal memory usage and allowing the memory concerns of the switch to be negligible. In contrast to hop-by-hop multipath, end-to-end multipath offers superior traffic distribution across multiple paths, particularly in asymmetric topologies \cite{b22}.

\begin{figure}[t]
	\centerline{\includegraphics[width=6.5cm]{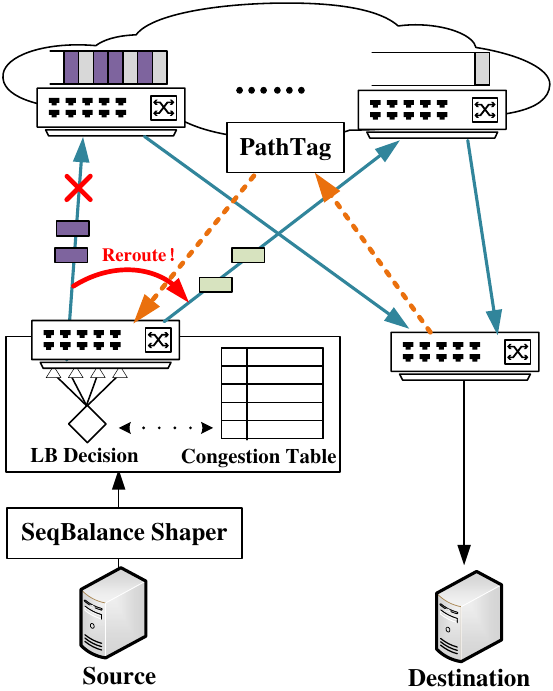}}
	\caption{Overview of SeqBalance.}
	\label{overview}
\end{figure}

\section{Design}
Avoiding out-of-order delivery of packets is especially important in RDMA transmissions. Therefore, we design SeqBalance, a new load balancing framework, which combines end-host fine-grained sub-flow division and in-network routing decision to provide a fine-grained load balancing mechanism.

\subsection{Overview of SeqBalance}\label{AA}
The key idea of SeqBalance is to avoid out-of-order delivery of packets as fine-grained as possible. Fig. 2 shows the system diagram of SeqBalance. Most of the function is located at the ToR switch. Load balancing decisions are primarily executed at the source ToR switch, relying on information stored in the Congestion Table. This table accumulates entries from the \textit{Congestion} Packet transmitted by the destination ToR switch. Notably, the path stored in the Congestion Table restricts the entry of new flows.At the end-host, integration of the \textit{SeqBalance Shaper} function module into the RDMA driver of RNIC achieves fine-grained flow control without necessitating other modifications.
\begin{figure*}[t]
	\begin{minipage}[t]{0.33\linewidth}
		\centering
		\includegraphics[height=3.3cm]{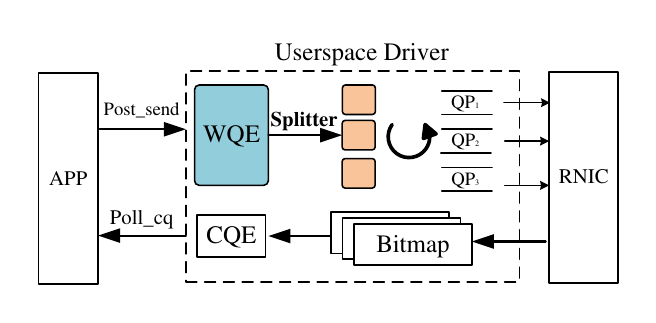}
		\caption{SeqBalance Shaper.}
		\label{fig:side:a}
	\end{minipage}%
	\begin{minipage}[t]{0.33\linewidth}
		\centering
		\includegraphics[height=4cm]{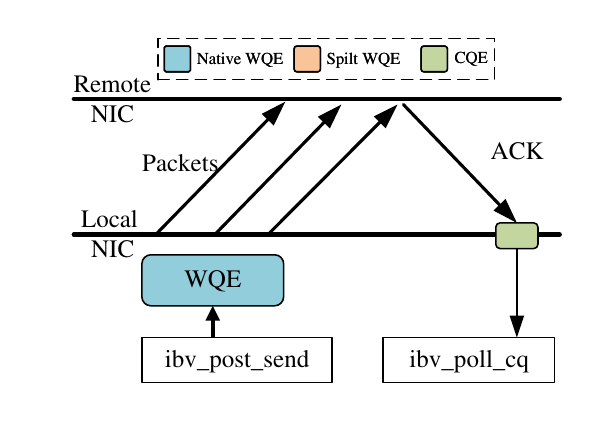}
		\caption{CQE Generation Scheme of RDMA.}
		\label{fig:side:b}
	\end{minipage}
	\begin{minipage}[t]{0.33\linewidth}
		\centering
		\includegraphics[height=3.5cm]{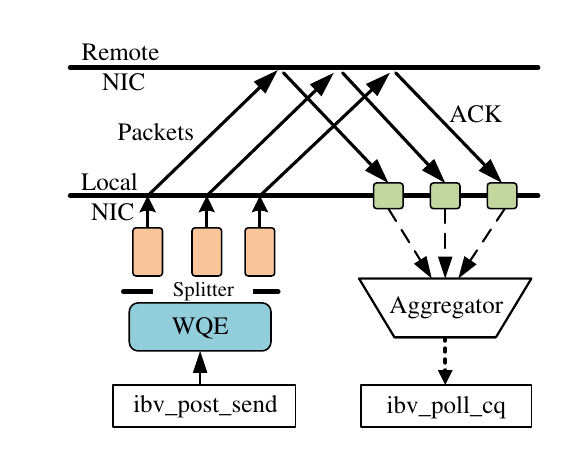}
		\caption{CQE Generation Scheme after SeqBalance Shaper.}
		\label{fig:side:b}
	\end{minipage}
\end{figure*}

The source ToR switch formulates load balancing decisions by analyzing the initial packet of each sub-flow. Path selection for subsequent packets is contingent upon their five-tuple, ensuring the continuity of using the same uplink. The destination ToR switch bears the responsibility of informing the source ToR switch about congested paths, designating them as bad paths, and updating the Congestion Table accordingly. Consequently, all load balancing decisions are made in-network without the need for end-host involvement.

\subsection{Routing Decisions}
Ideally, achieving per-packet level routing, as demonstrated by DRILL \cite{b12}, would maximize network utilization. However, adopting per-packet level routing introduces a substantial risk of out-of-order delivery for packets, significantly impacting the performance of RDMA transmission. Simultaneously, the flowlet-based routing mechanism proves unsuitable for RDMA transmission due to challenges in identifying flowlets within RDMA traffic. So how can we design a load balancing mechanism that is both fine-grained and can avoid out-of-order delivery?

Our routing mechanism only involves source ToR switches and destination ToR switches, ensuring a straightforward deployment. Upon the arrival of the initial packet of a sub-flow (refer to $\S$\uppercase\expandafter{\romannumeral3}.C for sub-flow division details) at the source ToR switch, it undergoes hashing based on its five-tuple, determining a specific path. If the chosen path lacks congestion markings, all ensuing packets of the sub-flow traverse through this designated path. Conversely, if the path bears congestion markings, the hashing process is reiterated until an uncongested path is selected. It is noteworthy that only the first packet of the sub-flow undergoes hashing, with subsequent packets adhering to the established transmission path. This is because RDMA implements the go-back-N mechanism. Once the transmission path is changed in the sub-flow and the subsequent packet arrives first, retransmission will be enabled, which will greatly affect the performance. Below are the functions performed by the source ToR switch and the destination ToR switch in the routing decision.
\\[4pt]
\textbf{Source ToR Switch. }The source ToR switch receives the \textit{Congestion} Packet from the destination ToR switch. Upon receipt of a \textit{Congestion} Packet, the source ToR switch designates the associated path as inactive for a fixed duration, denoted as $\phi$. Subsequent receptions of a \textit{Congestion} Packet marking the same path refresh the duration, restarting the timing from $\phi$. The path retains its inactive status until no \textit{Congestion} Packet is received within the specified time $\phi$. When a sub-flow is hashed to a path marked with the \textit{inactive} tag, it undergoes double hashing, and the path exclusively transmits sub-flows that are yet to be completed. The determination of the $\phi$ value is discussed in detail in $\S$\uppercase\expandafter{\romannumeral4}.A.
\\[4pt]
\textbf{Destination ToR Switch. }The destination ToR switch receives packets from the path. Upon receiving a packet marked with Explicit Congestion Notification (ECN), indicating congestion flagged by the Spine switch during transmission, the destination ToR switch initiates the transmission of a \textit{Congestion} Packet to the source ToR switch, conveying information about the path.

\subsection{SeqBalance Shaper}
Given the constraints of RDMA transmission, where per-packet-level routing (prone to out-of-order delivery) and flowlet-based approaches are challenging to implement due to RDMA traffic characteristics, the question arises: How can fine-grained load balancing be accomplished?

To address this, we introduce \textit{SeqBalance Shaper}, a functional module designed for flow splitting. Fig. 3 illustrates the schematic diagram of this module. \textit{SeqBalance Shaper} is capable of subdividing a flow into $N$ sub-flows for transmission. In RDMA, WQE is considered a "task description" containing detailed task information (e.g., sending data with a length of 10 bits at address 0x12345678 to the opposite node). Consequently, all packets associated with a WQE are transmitted as a unified flow. Our approach involves achieving flow splitting by segmenting the WQE. We achieve this by dividing the flow into $N$ sub-flows through the subdivision of the WQE into $N$ sub-WQEs.

Nevertheless, RDMA's order-preserving transmission relies on PSN bits, and a shared QP connection utilizes the same set of PSN. In other words, even if the WQE is fragmented into $N$ sub-WQEs and transmitted through the same QP, when these sub-flows traverse the source ToR switch and are subsequently hashed to different paths for transmission, the issue of out-of-order delivery arises. The cause of this problem lies in the order-preserving transmission within the same QP. To address this, we mitigate the out-of-order transmission problem by directing sub-WQEs through distinct QPs. At the same time, this design also corresponds to the business traffic of AI training. The number of streams trained by AI is small and the entropy is low. At this time, appropriately splitting WQE into different QP transmissions will increase the entropy multiple times, greatly improving the randomness of AI training traffic and thus improving load balancing efficiency.
However, QP represents a scarce resource in RNICs \cite{b17}, prompting a discussion on the trade-off involving the segmentation of the number of sub-WQEs in $\S$\uppercase\expandafter{\romannumeral3}.D.

\subsection{Implementation}

We implement SeqBalance’s load balancing decision on an Intel Tofino \cite{b19} programmable switch in $\sim$1000 lines of P4\_16\cite{b23} code. And we implement \textit{SeqBalance Shaper} in the RDMA driver with $\sim$1200 lines of C code.
\\[4pt]
\textbf{\textit{Congestion} Packet. }The destination ToR switch communicates path congestion to the source ToR switch through the transmission of a \textit{Congestion} Packet, temporarily restricting new flows from traversing the congested path. We designate 10 bits from the reserved field in the RDMA BTH as the PathTag field. This 10-bit PathTag field accommodates representation for 1023 paths within the topology, a sufficient number for conventional network topologies. For larger-scale network topologies, scalability is achievable by expanding the PathTag field's bit count. Upon the passage of a packet carrying an ECN tag through the destination ToR switch, the switch mirrors the packet, adjusts the PathTag field to include path information, and subsequently transmits it back to the source ToR switch to signal path congestion. The source ToR switch deciphers the path congestion information from the packet header, marking it as inactive for a duration of $\phi$.
\\[4pt]
\textbf{Completion Queue Element (CQE). }In RDMA, CQE corresponds to WQE and can be perceived as a "task report" returned to the software following hardware task completion. Fig. 4 illustrates the original RDMA's CQE generation scheme. Upon task completion at the receiver end, an ACK is transmitted to the sender end, indicating task fulfillment. Subsequently, upon ACK reception, the sender generates a CQE to notify the application of task completion. However, \textit{SeqBalance Shaper} introduces segmentation of WQE, leading to concerns about CQE generation. The upper-layer application dispatches the complete WQE without awareness of \textit{SeqBalance Shaper}'s WQE segmentation, necessitating a comprehensive CQE. Consequently, addressing the challenge of generating a complete CQE after WQE segmentation becomes crucial.

Currently, two viable solutions exist: the first approach resembles the initial CQE generation scheme of RDMA, involving the transmission of a comprehensive ACK after receiving all sub-WQEs at the receiver, resulting in the generation of a complete CQE. The second solution entails each sub-WQE independently generating its ACK, and upon receiving all ACKs, the sender produces a complete CQE to inform the upper-layer application of task completion.

We have opted for the second solution to implement CQE. This choice is informed by the limitation of the first solution, where the receiver lacks awareness of WQE segmentation. In the first solution, the receiver is only tasked with performing operations upon the arrival of the WQE, without knowledge of whether the WQE is a standalone or a sub-WQE of a complete WQE. This lack of information makes the implementation of this scheme challenging. Conversely, the second solution allows us to ascertain the completion of all sub-WQEs at the sender. In other words, upon receiving ACKs for all sub-WQEs, the sender can then dispatch CQEs to the upper-layer application. Fig. 5 illustrates our CQE generation scheme, where each segmented WQE independently generates its ACK. The complete CQE is generated only after receiving all ACKs, signaling completion to the upper-layer application.

So how can we ensure that the ACKs of all sub-WQEs have been received? To address this, we have chosen to employ the \textit{Bitmap} data structure for implementation. \textit{Bitmaps} are utilized to keep track of received ACKs for sub-WQEs. In contrast to the receiver's lack of information about WQE details, the sender possesses knowledge of the number of sub-WQEs the WQE is divided into, the specific QP each sub-WQE is transmitted through, and related information. Consequently, employing the \textit{Bitmap} data structure allows for a straightforward realization of monitoring ACKs for all sub-WQEs. Subsequently, upon receipt of all ACKs for sub-WQEs, a CQE is transmitted to notify the upper layer that the task has been completed.
\\[4pt]
\textbf{SeqBalance Shaper. }We have integrated \textit{SeqBalance Shaper} into the RDMA driver. As detailed in $\S$\uppercase\expandafter{\romannumeral3}.C, \textit{SeqBalance Shaper} segments the WQE, thereby dividing the flow into sub-flows to achieve fine-grained load balancing. To prevent out-of-order packet delivery, \textit{SeqBalance Shaper} allocates the segmented sub-WQEs to different QPs for transmission. However, exceeding a certain threshold in the number of QPs can lead to frequent context switching, resulting in cache misses and subsequent performance degradation (refer to $\S$\uppercase\expandafter{\romannumeral5} for an in-depth discussion). Therefore, achieving a balance between the number of divided sub-WQEs (i.e., the number of QPs), denoted as $N$, and load balancing performance is crucial. For businesses like AI training with fewer flows, increasing the number of $N$ can improve the performance of load balancing; If it is a business with a large number of flows, the number of $N$ can be reduced (in extreme cases, it can be taken as 1) to ensure the scalability of QP.

In Fig. 6, we illustrate the variation in FCT with $N$, ranging from 2 to 6. It is evident that as $N$ increases, both the average FCT and $99^{th}$ percentile FCT decrease. Specifically, when $N$ = 4, compared to $N$ = 2, the average FCT and $99^{th}$ percentile FCT show reductions of 25.4\% and 27.6\%, respectively. With $N$ = 6, compared to $N$ = 4, the reductions are 15.1\% and 14.9\%, respectively, but this comes at the cost of a 50\% increase in QP resource consumption. Fig. 7 focuses on load balancing efficiency for different $N$, presenting the Cumulative Distribution Function (CDF) of throughput imbalance across the 8 uplinks for each ToR switch. Throughput imbalance\cite{b6} is defined as the maximum throughput (among the 8 uplinks) minus the minimum, divided by the average. We observe that when $N$ \textgreater 4, the improvement in load balancing efficiency diminishes. Therefore, to strike a balance between $N$ performance and load balancing, we opt for a compromise value of $N$ = 4 in subsequent large-scale simulations and testbed experiments.

\section{Evaluation}
In this section, we use testbed experiments with our prototype and large-scale NS3 simulations \cite{b26} to evaluate the performance of SeqBalance and compare with the-state-of-art load balancing schemes.

\begin{figure}[t]
	\centering
	\begin{tabular}[c]{@{}r@{}l@{}}
		\subfigure[Avg FCT] 
		{
			
			\centering          
			\includegraphics[width=4cm]{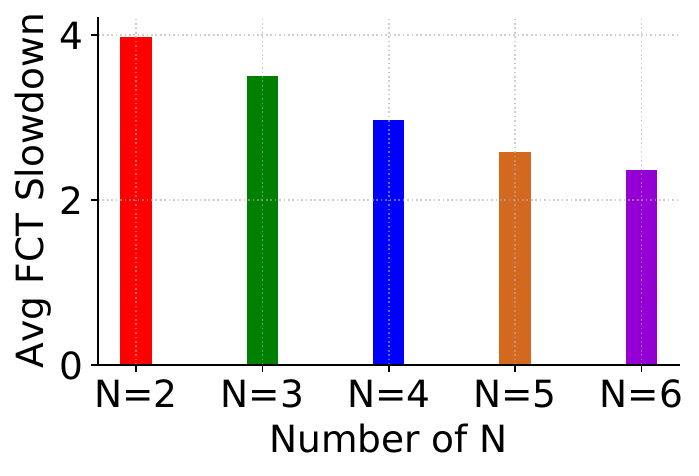}   
			
		}
		\subfigure[$99^{th}$-ile FCT] 
		{
			
			\centering          
			\includegraphics[width=4cm]{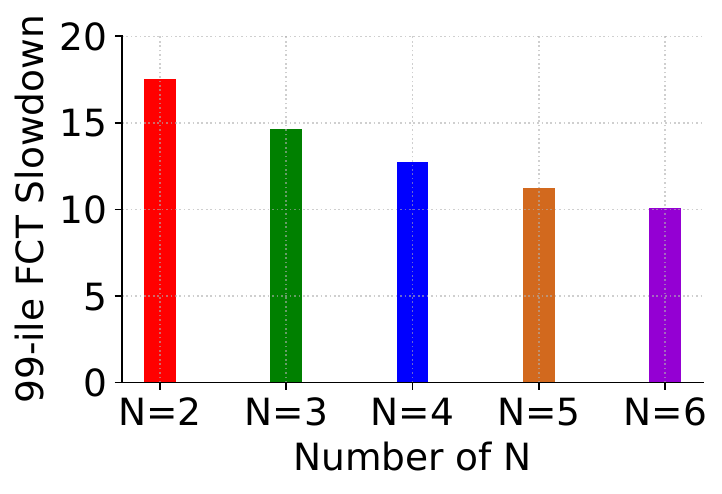}   
			
		}
	\end{tabular}
	\caption{FCT comparison with different N.}
\end{figure}

\begin{figure}[t]
	\centerline{\includegraphics[width=7cm]{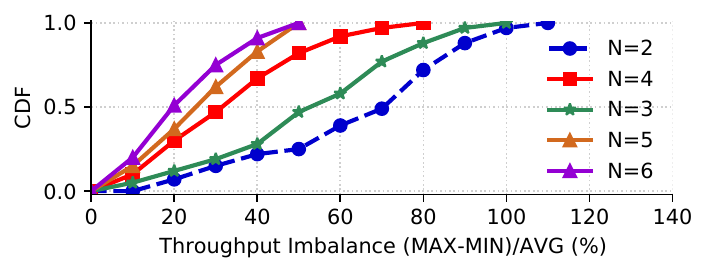}}
	\caption{Extent of imbalance between throughput of uplinks with different N.}
	\label{Traffic}
\end{figure}

\begin{figure}[t]
	\centering
	\begin{tabular}[c]{@{}r@{}l@{}}
		\subfigure[Symmetric topology.]
		{
			
			\centering          
			\includegraphics[width=4.5cm]{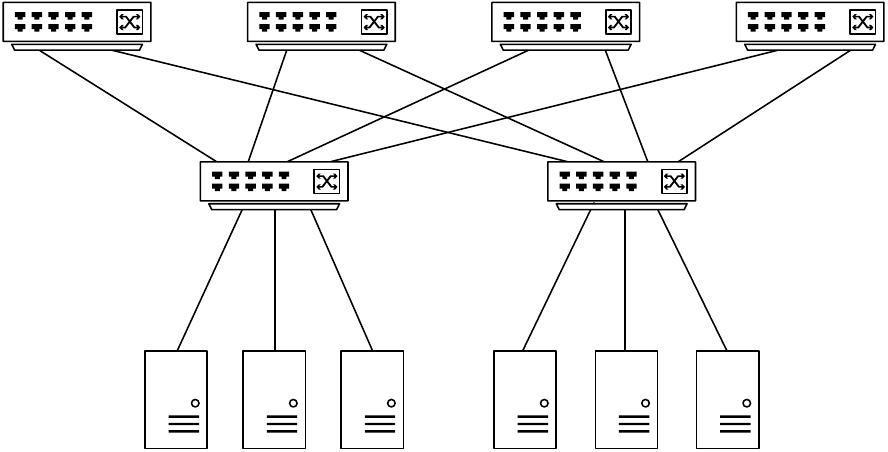}   
			
		}
		\subfigure[Asymmetric topology.] 
		{
			
			\centering          
			\includegraphics[width=3.3cm]{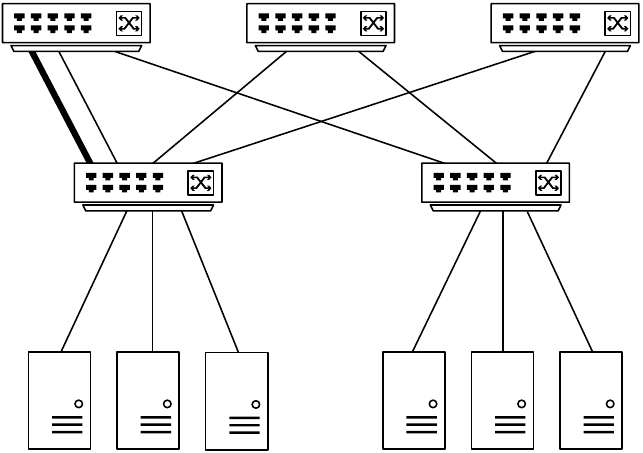}   
			
		}
	\end{tabular}
	\caption{Topologies in testbed.}
\end{figure}

\begin{figure}[t]
	\centerline{\includegraphics[width=7cm]{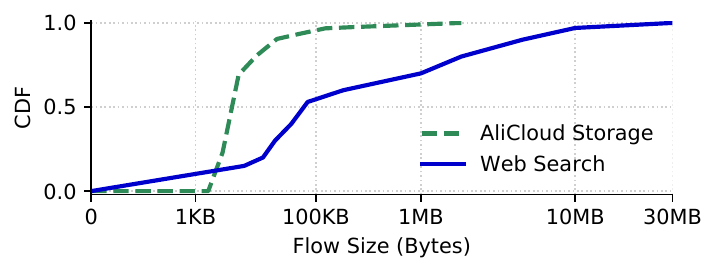}}
	\caption{Traffic load distribution from AliCloud Storage\cite{b27} and WebSearch\cite{b4}.}
	\label{Traffic}
\end{figure}

\subsection{Testbed Experiments}
We first introduce the setup of the testbed experiments.
\\[4pt]
\textbf{Network topology. }By default, we employ the leaf-spine topology commonly observed in data center networks. The testbed topology comprises 2 leaf switches and 4 spine switches, as illustrated in Fig. 8 (a). Specifically, we utilize Tofino switches for the two leaf switches, while the four spine switches employ commodity Ethernet switches. Notably, each leaf switch is interconnected with 3 servers, and each server is equipped with a CX6 RNIC \cite{b18}. All network links are configured with a bandwidth of 40Gbps.
\\[4pt]
\textbf{Schemes compared. }We conduct a comparative analysis between SeqBalance and ECMP. In the case of SeqBalance, the parameter $\phi$ is configured based on the ECN marking threshold, where $\phi$ corresponds to the time required to refresh the threshold for the link rate. Considering our testbed's 40Gbps link, and following the ECN threshold recommendation of 160KB from \cite{b27}, we set the value of $\phi$ to 32$us$.
\\[4pt]
\textbf{Traffic loads. }We employ Perftest\cite{b33} to establish enduring connections among three client-server pairs. For bandwidth-intensive applications, we utilize \textit{ib\_write\_bw}. This application employs 4 QPs, with each QP maintaining 128 outstanding 2MB WRITE requests to fully utilize the link bandwidth.
\\[4pt]
\textbf{Flow congestion control. }SeqBalance is a load balancing scheme for RoCE. For the traffic congestion control (CC) scheme of RNIC, we use DCQCN \cite{b4}. For the parameter configuration of DCQCN, we refer to the observations in \cite{b27} and use the following parameters $(K_{min}, K_{max}, P_{max})$ = (160KB, 520KB, 0.2). Additionally, all switches are configured with PFC. The retransmission strategy implemented on the RNIC is Go-Back-N.
\\[4pt]
\textbf{Performance metrics. }We use port rate as the main performance metrics. For port rate, we measure the ingress rate on the switches.
\\[5pt]
\textbf{Load balancing efficiency. }To assess load balancing efficiency, we gauge the port rates of switches along the four paths to depict the load distribution. We activate a server every 150$s$ to initiate traffic. In Fig. 10, the alterations in port rates for both schemes are illustrated over a 450$s$ period. The maximum load can reach 75\% when all three servers are operational. Fig. 10(a) details the path rates under ECMP. In the initial 150$s$, only one server injects traffic, involving three paths. From 150$s$ to 300$s$, as the second server contributes traffic, the bandwidth of the third path saturates. Subsequently, between 300$s$ and 450$s$, with the third server adding traffic, the total rate across the three links only reaches 101.4Gbps. Hash collisions lead to load imbalance, triggering a decrease in CC rates. In contrast, SeqBalance achieves a relatively balanced load through finer-grained multi-path transmission and an in-network load balancing strategy. This effective approach avoids the rate decrease associated with CC activation.
\\[4pt]
\textbf{Bandwidth overhead of \textit{Congestion} Packet. }In this subsection, we examine the bandwidth overhead incurred by \textit{Congestion} Packets. SeqBalance dispatches \textit{Congestion} Packets to the source ToR switch through the destination ToR switch to convey path congestion information. Bandwidth overhead is assessed by tracking the total count of \textit{Congestion} Packets at the source ToR switch within a specific timeframe. Table \uppercase\expandafter{\romannumeral2} presents the bandwidth overhead of \textit{Congestion} Packets for three different loads. Observations reveal that at a 25\% load, no \textit{Congestion} Packets are generated; at a 50\% load, the bandwidth consumed by \textit{Congestion} Packets is merely 4Kbps, which is nearly negligible; and at a 75\% load, the bandwidth utilized by \textit{Congestion} Packets is 0.05Gbps. This indicates that \textit{Congestion} Packets occupy only a small fraction of the bandwidth compared to data transmission. Consequently, the presence of \textit{Congestion} Packets does not impede normal data transmission.

\begin{figure}[t]
	\centering
	\begin{tabular}[c]{@{}r@{}l@{}}
		\subfigure[SeqBalance] 
		{
			
			\centering          
			\includegraphics[width=4cm]{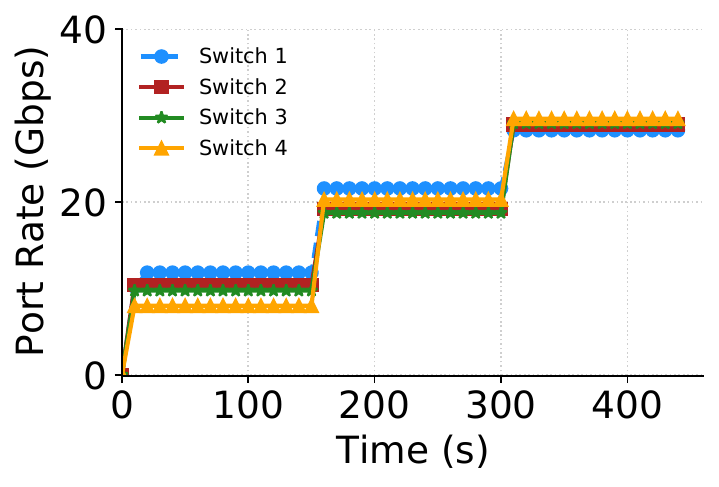}   
			
		}
		\subfigure[ECMP] 
		{
			
			\centering          
			\includegraphics[width=4cm]{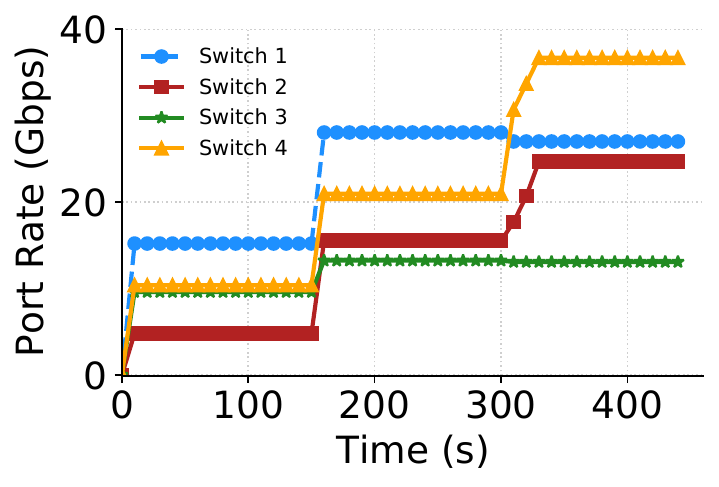}   
			
		}
	\end{tabular}
	\caption{The port rate of the switches in symmetric topology.}
\end{figure}
\begin{figure}[t]
	\centering
	\begin{tabular}[c]{@{}r@{}l@{}}
		\subfigure[SeqBalance] 
		{
			
			\centering          
			\includegraphics[width=4cm]{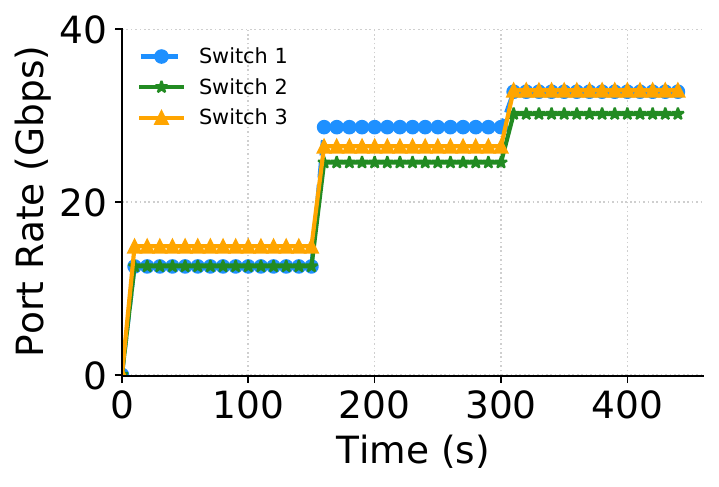}   
			
		}
		\subfigure[ECMP] 
		{
			
			\centering          
			\includegraphics[width=4cm]{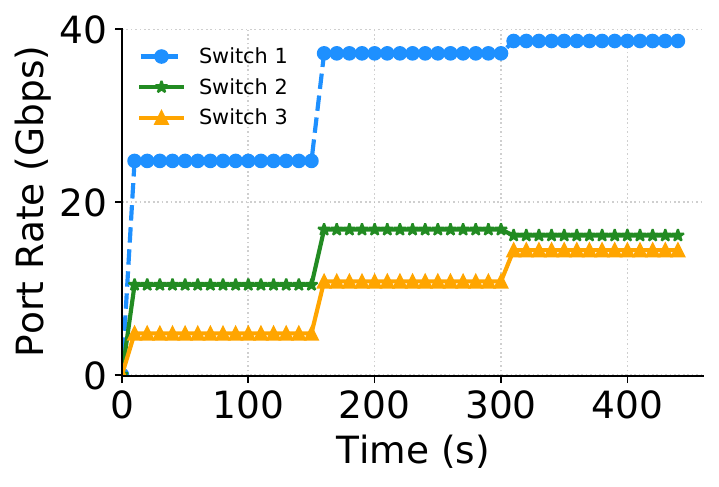}   
			
		}
	\end{tabular}
	\caption{The port rate of the switches in asymmetric topology.}
\end{figure}
\begin{figure*}[t]
	\centering
	\begin{tabular}[c]{@{}r@{}l@{}}
		\subfigure[Avg FCT for the AliCloud Storage workload.] 
		{
			
			\centering          
			\includegraphics[width=4cm]{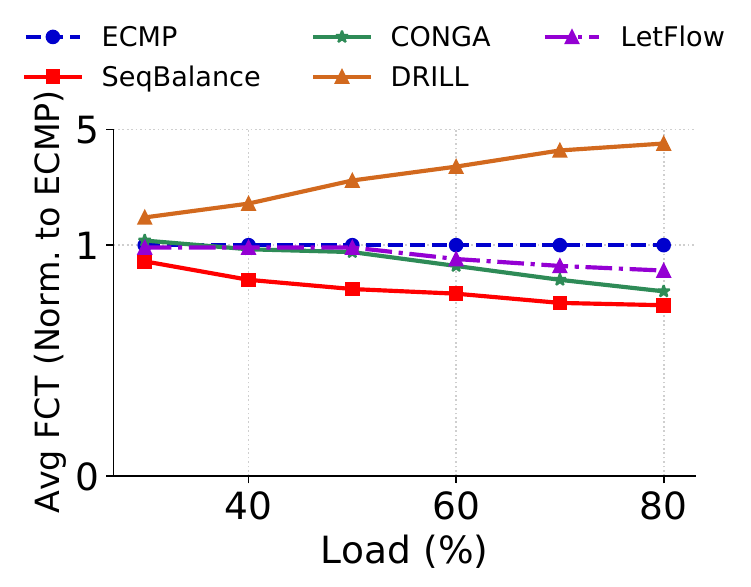}   
			
		}
		\subfigure[$99^{th}$-ile FCT for the AliCloud Storage workload.] 
		{
			
			\centering          
			\includegraphics[width=4cm]{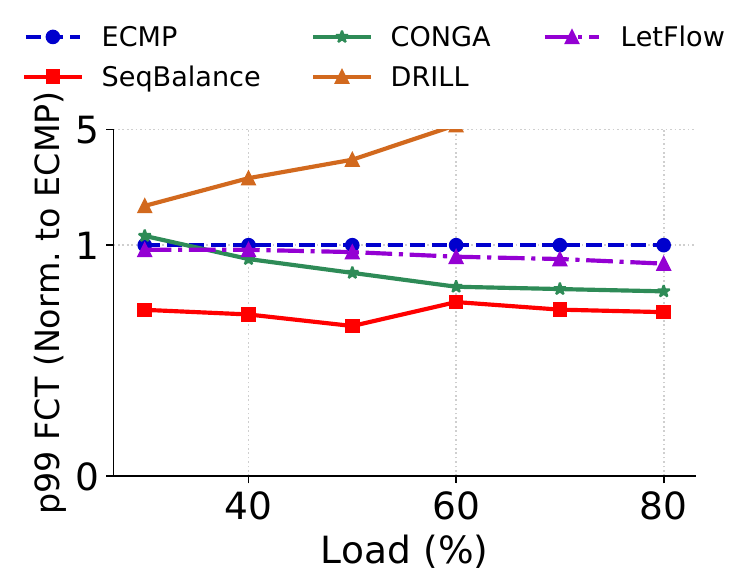}   
			
		}
		\subfigure[Avg FCT for the WebSearch workload.] 
		{
			
			\centering          
			\includegraphics[width=4cm]{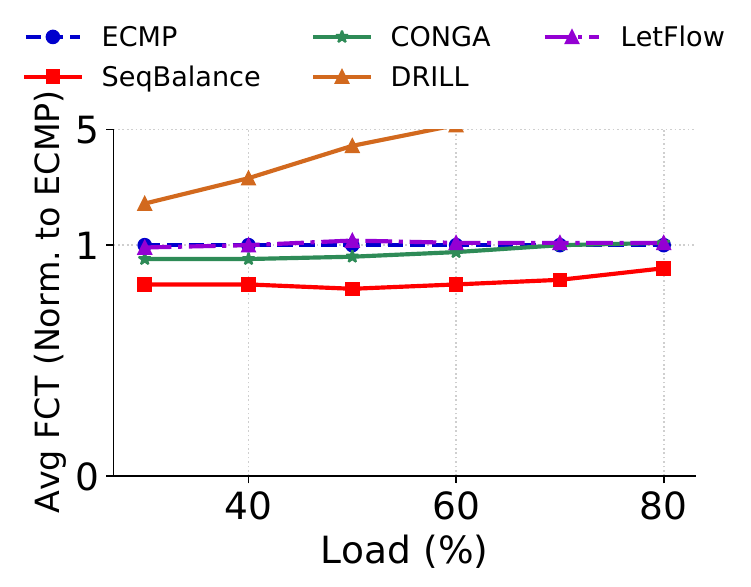}   
			
		}
		\subfigure[$99^{th}$-ile FCT for the WebSearch workload.] 
		{
			
			\centering          
			\includegraphics[width=4cm]{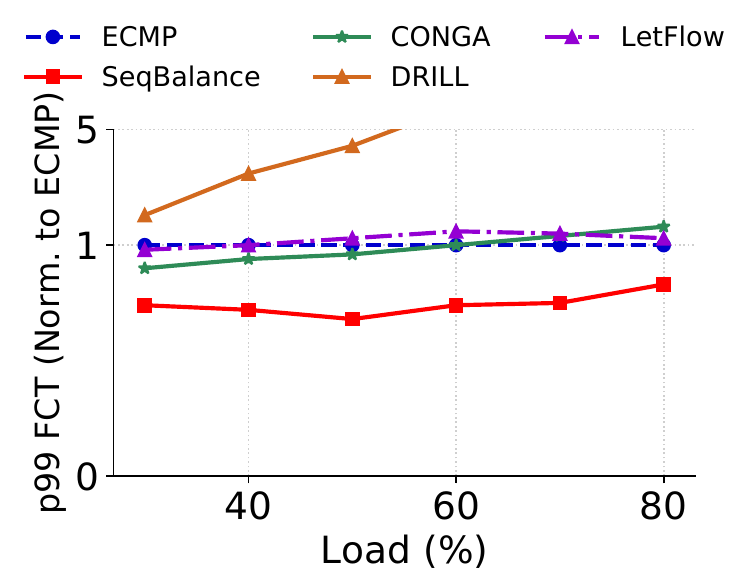}   
			
		}
	\end{tabular}
	\caption{FCT comparison for two workloads.}
\end{figure*}
\noindent\textbf{Asymmetric topology. }In this subsection, we examine the performance of SeqBalance in the asymmetric topology. Fig. 8 (b) illustrates the architectural diagram of the asymmetric topology within our testbed. In this setup, one of the switches is deactivated, and the link is redirected to the other switch, resulting in one of the three uplinks of the ToR switch operating at 80Gbps, while all other links remain at 40Gbps. Other configurations remain consistent with those of the symmetric topology. Fig. 11 depicts the load balancing efficiency of both schemes on the three paths over a 450$s$ duration. A notable observation is that, in the asymmetric topology, ECMP exhibits lower total rates within the 300-450$s$ interval compared to the symmetric topology. This discrepancy arises because ECMP, being a hop-by-hop load balancing scheme, lacks awareness of downstream link conditions. Consequently, load imbalance triggered by hash collisions induces a rate reduction in the CC mechanism. Fig. 11 (a) illustrates the port rates of SeqBalance's switches. Under the asymmetric topology, SeqBalance demonstrates a relatively balanced load. In contrast to ECMP, SeqBalance achieves a 37.6\% higher throughput. This improvement is attributed to SeqBalance's implementation of an end-to-end routing strategy, enabling effective handling of asymmetric topologies.

\begin{table}[t]
	\begin{center}
		\caption{Bandwidth overhead of Congestion Packet.}
		\begin{tabular}{c|cc} 
			\multirow{2}{*}{\textbf{Load}} & \multicolumn{2}{c}{\textbf{Bandwidth}}\\
			\cline{2-3}
			& \textbf{RDMA Packet} & \textbf{\textit{Congestion} Packet} \\
			\hline
			25 & 39.81Gbps & 0.00bps\\
			50 & 79.54Gbps & 4.00Kbps\\
			75 & 115.94Gbps & 0.05Gbps\\
		\end{tabular}
	\end{center}
\end{table}
\begin{figure}[t]
	\centering
	
	\subfigure[ALiCloud Storage workload] 
	{
		
		\centering          
		\includegraphics[width=7cm]{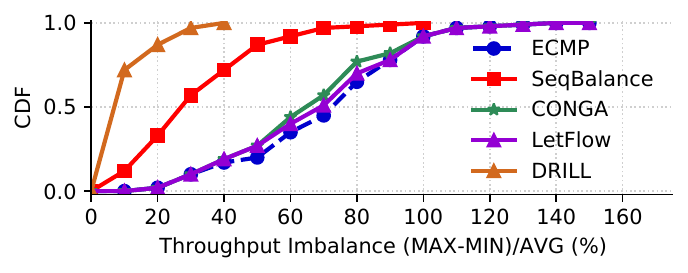}   
		
	}
	\subfigure[WebSearch workload] 
	{
		
		\centering          
		\includegraphics[width=7cm]{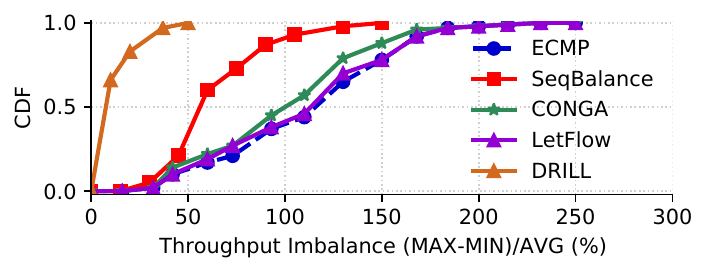}   
		
	}
	
	\caption{Extent of imbalance between throughput of leaf up-links for both workloads at 80\% load.}
\end{figure}
\subsection{Large Scale Simulation}
\begin{figure*}[t]
	\centering
	\begin{tabular}[c]{@{}r@{}l@{}}
		\subfigure[Avg FCT for AliCloud Storage workload.] 
		{
			
			\centering          
			\includegraphics[width=4cm]{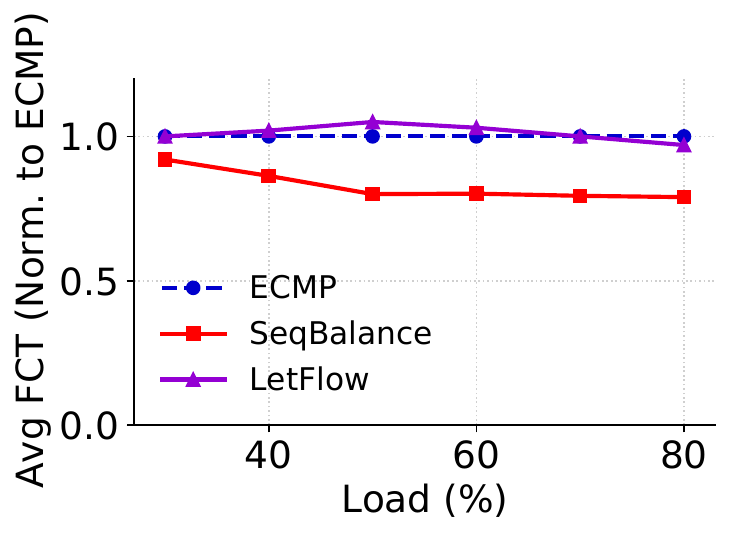}   
			
		}
		\subfigure[$99^{th}$-ile FCT for AliCloud Storage workload.] 
		{
			
			\centering          
			\includegraphics[width=4cm]{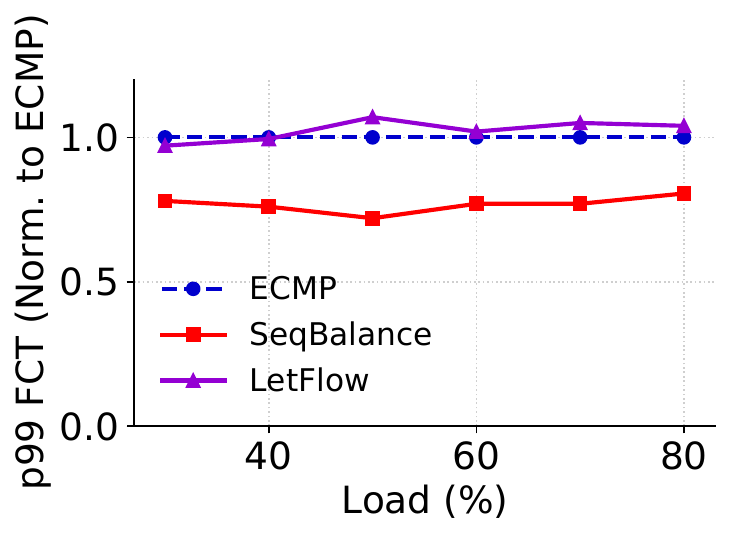}   
			
		}
		\subfigure[Avg FCT for WebSearch workload.] 
		{
			
			\centering          
			\includegraphics[width=4cm]{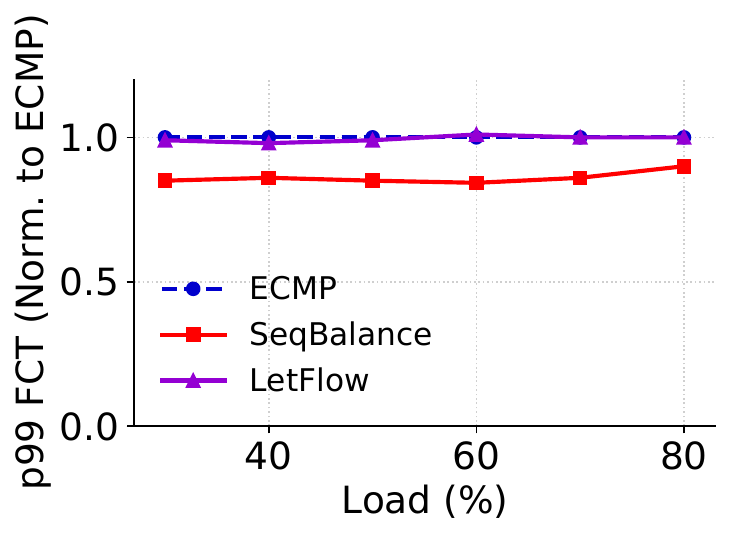}   
			
		}
		\subfigure[$99^{th}$-ile FCT for WebSearch workload.] 
		{
			
			\centering          
			\includegraphics[width=4cm]{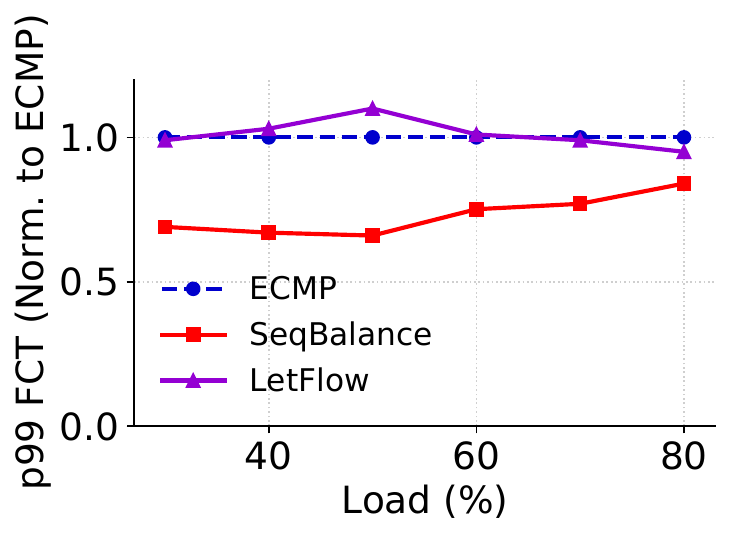}   
			
		}
	\end{tabular}
	\caption{FCT comparison for two workloads with 3-tier topology.}
\end{figure*}
Next, we evaluate SeqBalance in large-scale NS3 simulation.
\\[4pt]
\textbf{Network topology. }In line with the testbed experiments, our simulation adopts a leaf-spine topology. The simulated topology comprises 8 leaf switches and 12 spine switches, with each leaf switch connected to 16 servers, totaling 128 servers. All links operate at a rate of 100Gbps, and the delay is set to 1$us$.
\\[4pt]
\textbf{Schemes compared. }We conduct a comparative analysis involving SeqBalance, ECMP, LetFlow\cite{b14}, CONGA \cite{b6} and DRILL\cite{b12}—state-of-the-art load balancing schemes for data center networks. Regarding the flowlet parameters of CONGA, we opt for a 100$us$ inactive time gap threshold. In the case of SeqBalance, given the variability in link rates and ECN thresholds, we configure the value of $\phi$ to be 32$us$.
\\[4pt]
\textbf{Traffic loads. }We employ the ALiCloud Storage and Web Search \cite{b6} workloads, with the load distribution depicted in Fig. 9. ALiCloud Storage traffic is predominantly concentrated in the 1KB-100KB range, primarily comprising small flows. In contrast, Web Search traffic exhibits a mix of small flows ($\textless$100KB) and large flows (\textgreater10MB). During flow generation, we randomly designate two out of the 128 servers as the client and server to initiate a flow, selecting the flow size based on the load distribution.
\\[4pt]
\textbf{Flow congestion control. }We employ the DCQCN as the CC scheme in NS3 simulations. Given the simulation's link rate of 100Gbps, we set the parameters $(K_{min}, K_{max}, P_{max})$ to (400KB, 1600KB, 0.2), following the recommendations in \cite{b27}. Additionally, switches are configured with PFC.
\\[4pt]
\textbf{Performance metrics. }We employ FCT slowdown and throughput imbalance as the primary performance metrics. FCT slowdown represents the normalization of actual FCT to the optimal FCT achievable in an idle network.
\\[5pt]
\textbf{Impact on FCT. }We conduct simulations across traffic loads ranging from 30\% to 80\%. In Fig. 12, we present the results for two distinct workloads. Parts (a) and (c) display the average FCT for each scheme, while parts (b) and (d) depict the $99^{th}$-percentile FCTs for each scheme. All values are normalized to those achieved by ECMP.

We observe that under the AliCloud Storage workload, compared to other algorithms, SeqBalance reduced the average FCT by 16\% and the $99^{th}$ FCT by 23.6\%. Under WebSearch load, SeqBalance reduced the average FCT by 13.5\% and the $99^{th}$ FCT by 27.6\%. This is because ECMP is flow granularity, resulting in the worst performance due to hash conflicts. For CONGA and LetFlow, the performance is not as good as that used in TCP because it is almost difficult to detect flowlets in RDMA. DRILL triggers retransmission due to packet granularity, so FCT is much higher than the other four algorithms. At the same time, some data is not displayed in the graph due to being too high. SeqBalance achieves better load balancing performance through flow splitting and congestion aware routing.
\\[4pt]
\textbf{Load balancing efficiency. }In this subsection, we delve into the load balancing efficiency of SeqBalance. Fig. 13 illustrates the CDF of throughput imbalance across the uplinks for each ToR switch, specifically at the 80\% load level for both workloads. We employ sampling at ToR switch ports every 100 $us$ to calculate throughput. The results reveal that, for both workloads, SeqBalance significantly outperforms ECMP and CONGA, except for DRILL. However, DRILL, relying on per-packet load balancing, resorts to packet spraying, inducing out-of-order packets and a sharp increase in FCT.
\\[4pt]
\textbf{Impact on FCT with 3-tier topology. }The preceding simulations were conducted on a 2-tier topology. In this subsection, we assess the performance of SeqBalance within a 3-tier topology. A 3-tier topology involves more extended end-to-end paths between servers, leading to prolonged response times. The 3-tier topology chosen is FatTree\cite{b29}, comprising 16 core switches, 20 Agg switches, and 20 ToR switches. Each ToR switch is linked to 16 servers, totaling 320 servers. The link speed connecting the Agg switch and the ToR switch is 400Gbps, while all other links operate at 100Gbps. Consistent with \cite{b27}, all link delays are set to 1 $us$. 
Other configurations mirror those employed in the 2-tier topology.

We excluded CONGA from consideration in the 3-tier topology due to its design tailored for 2-tier topology \cite{b6}. Our comparative analysis involves SeqBalance, ECMP, and LetFlow. Fig. 14 presents the FCT results for two workloads. It can be seen that in a larger scale 3-layer topology, SeqBalance also maintains its performance advantage. Under the ALiCloud Storage workload, SeqBalance reduced the average FCT by 18.7\% and the $99^{th}$ FCT by 27.1\%. Under WebSearch load, SeqBalance reduced the average FCT by 13.1\% and the $99^{th}$ FCT by 33.2\%.



\section{Discussion}
\textbf{Adapting to larger networks. }Given the finite hardware resources in switches and RNICs, challenges emerge as the network scales up. The switch's resource demand pertains to the storage memory allocated for the \textit{Congestion Table}, contingent upon the count of congested paths. In extensive cross-pod networks, the number of congested paths remains within the limits of network topology connections. Consequently, the switch can adequately accommodate congested path information even in scenarios where all paths experience congestion. The RNIC's QP resources are contingent on the connections established by upper-level applications, showing no significant surge with network size escalation. Under AI training traffic, the characteristic of having fewer flows perfectly fits SeqBalance.
Additionally, \textit{SeqBalance Shaper} demonstrates the capability to adjust the number of sub-WQEs in accordance with network size variations. Thus, SeqBalance exhibits adaptability to larger networks.
\\[4pt]
\textbf{Co-design with connection scalability. }RNICs have always had connection scalability issues. The challenge arises when the number of QPs surpasses a particular threshold, resulting in a substantial decline in RDMA performance \cite{b17, b30, b31, b32}. Although the inner workings of commercial RNICs are opaque, the fundamental cause of this performance deterioration is attributed to cache misses stemming from context switches between connections \cite{b30}. In essence, QP emerges as a tightly constrained resource within RNIC. \textit{SeqBalance Shaper} introduces WQE segmentation, breaking down a WQE into $N$ sub-WQEs for transmission. Undoubtedly, this approach moderately escalates the consumption of QP resources. However, it is crucial to note that for scenarios involving a multitude of flows transmitted through a single RNIC, a load balancing framework like SeqBalance may not be the most suitable solution. Exploring the co-design of SeqBalance and research on connection scalability, as presented in \cite{b17}, promises to be an intriguing avenue for future investigation.
\\[4pt]
\textbf{Co-design with more congestion signals. }SeqBalance adopts an in-network routing strategy to avoid congested paths. The design of congestion perception utilizes the ECN signal of DCQCN\cite{b4}. The congestion perception design based on ECN signals has a significant advantage, which is that DCQCN is the most widely used CC algorithm in existing RNICs. The ECN signals from DCQCN can be easily integrated into existing networks without the need to build new protocols. However, other congestion signals such as sensing link load and RTT on switches require the construction of new protocols to support them, which increases the difficulty of large-scale deployment. Of course, the use of more congestion signals may increase the accuracy of congestion perception. Therefore, how to balance the difficulty of large-scale deployment and the accuracy of congestion perception is a question worth studying in the future.

\section{Related Work}
\textbf{Load balancing for TCP. }In the realm of load balancing, the foundational work is represented by ECMP. However, ECMP, rooted in flow granularity, can result in load imbalance, especially in scenarios with an uneven distribution of large and small flows within the data center. 
Consequently, much research on load balancing in data center networks has shifted toward more refined mechanisms. Flowlet-based load balancing stands out as a significant direction in this pursuit \cite{b6, b14}. CONGA \cite{b6} dissects traffic into flowlets, utilizing an in-network congestion feedback mechanism to achieve load balancing. The implementation of LetFlow \cite{b14} is more straightforward; it randomly assigns a path for each flowlet, with flowlets dynamically resizing based on path congestion. However, the applicability of flowlet-based schemes to RDMA scenarios is limited due to the characteristics of RDMA traffic, which hinders the identification of flowlets in RDMA flows compared to TCP flows. In contrast, SeqBalance distinguishes itself by not relying on the discovery of flowlets for ensuring in-order delivery. It adeptly circumvents out-of-order delivery issues through a thoughtful design.

In addition to flowlet-based load balancing, another prominent approach in load balancing for data centers is per-packet-based load balancing. The per-packet strategy, characterized by fine-grained transmission, facilitates improved decision-making, leading to enhanced overall performance. DRILL \cite{b12} distinguishes itself by implementing rerouting on a per-packet basis, resulting in nearly optimal load balancing. Despite DRILL's efforts to mitigate out-of-order packet delivery through thoughtful design, its applicability to RDMA scenarios is limited \cite{b34}. Presto \cite{b7} shares a similar concept but classifies all flows into 64KB "flowcells", slightly larger than per-packet-based load balancing. While this method attempts to minimize out-of-order packet delivery, it cannot entirely eliminate it, leading to suboptimal performance in RDMA transmissions.
In contrast, while SeqBalance does not achieve per-packet granularity but rather sub-flow granularity, it effectively avoids out-of-order delivery, thereby mitigating the performance impact of out-of-order packets on RDMA transmissions.
\\[5pt]
\textbf{Load balancing for RDMA. }There are presently several studies focusing on RDMA load balancing \cite{b24, b39, b40}. MP-RDMA \cite{b16} achieves multi-path RDMA transmission through FPGA hardware. However, RNICs based on custom designs lack compatibility with those commonly found in today's data centers. 
Maestro \cite{b24} introduces a multi-path load balancing mechanism by modifying the \textit{librdmacm} and \textit{libibverbs} libraries. This approach virtualizes multiple vNICs to split flows and transmits flow tables to switches through the SDN controller. Nevertheless, it relies on SDN controller support, requiring comprehensive reconfiguration when paths are altered.
ConWeave\cite{b34} presents an in-network load balancing scheme for RDMA, utilizing the queue pause/resume function of the Tofino2 programmable switch in its design. 
ConWeave implements rerouting through switches and leverages two queues on the destination ToR switch to reorder a single flow. However, the effectiveness of ConWeave's reordering is contingent on the number of queues on the switch. 
Presently, even the most advanced Tofino2 programmable switch is equipped with only 128 queues, rendering ConWeave unsuitable for scenarios characterized by substantial traffic.
Proteus\cite{b40} conduct in-depth research on the impact of PFC on RDMA load balancing and propose a PFC-aware load balancing scheme that is resilient to PFC pausing by exploring a combination of multi-level congestion signals. 
However, this solution require all switches to support programmability, and the deployment cost is too high.

In contrast, the implementation of SeqBalance only requires the addition of the \textit{SeqBalance Shaper} module to the RDMA software driver on the end-host side to enable fine-grained transmission. SeqBalance accomplishes congestion awareness and routing decisions within the network through programmable switches, overcoming limitations posed by the resources of programmable switches, such as queues and memory.
%

\section{Conclusion}
In this paper, we propose Seqbalance, a load balancing framework tailored for RDMA. Seqbalance adeptly achieves fine-grained multi-path load balancing while avoiding packet out-of-order, and effectively leveraging the diverse end-to-end paths within data centers. To accomplish nuanced RDMA routing without inducing packet out-of-order, we devise the \textit{Seqbalance Shaper}, a functional module embedded in the RNIC driver. This module divides a RDMA flow into $N$ sub-flows for transmission, thereby realizing RDMA multi-path transmission. Crucially, the upper-layer application remains oblivious to flow segmentation, ensuring that the delivery of WQE and upload of CQE by the upper-layer application remain unaffected. Simultaneously, we offload the routing decision to the ToR switch, enabling in-network load balancing. The Seqbalance prototype has been implemented on the Mellanox CX-6 RNIC and Tofino programmable switch, affirming the viability of Seqbalance's hardware implementation and compatibility with current commercial RNIC implementations. Evaluations conducted on our hardware testbed experiments and extensive simulations underscore Seqbalance's superior performance compared to state-of-the-art load balancing schemes for RoCE.

\bibliographystyle{unsrt}
\bibliography{RDMAreferences.bib}

\end{document}